\documentclass[twocolumn,english,aps,prb]{revtex4-1}
\usepackage{ae,aecompl}
\usepackage{helvet}

\usepackage[T2A,T1]{fontenc}
\usepackage[latin9]{inputenc}
\usepackage{graphicx}
\usepackage{esint}

\makeatletter
\@ifundefined{textcolor}{}
{%
 \definecolor{BLACK}{gray}{0}
 \definecolor{WHITE}{gray}{1}
 \definecolor{RED}{rgb}{1,0,0}
 \definecolor{GREEN}{rgb}{0,1,0}
 \definecolor{BLUE}{rgb}{0,0,1}
 \definecolor{CYAN}{cmyk}{1,0,0,0}
 \definecolor{MAGENTA}{cmyk}{0,1,0,0}
 \definecolor{YELLOW}{cmyk}{0,0,1,0}
}

\AtBeginDocument{\DeclareFontEncoding{T2A}{}{}}\usepackage{babel}

\usepackage{babel}

\usepackage{babel}

\usepackage{babel}

\usepackage{babel}

\makeatother

\usepackage{babel}
\begin{document}

\title{Superfluid density of a pseudogapped superconductor near SIT }

\author{M. V. Feigel'man$^{1,2}$ and L. B. Ioffe$^{3,1}$}

\affiliation{$^{1}$ L. D. Landau Institute for Theoretical Physics, Chernogolovka,
Moscow region, Russia}

\affiliation{$^{2}$ Moscow Institute of Physics and Technology, Dolgoprudny,
Moscow region, Russia}

\affiliation{$^{3}$ LPTHE, Universite Pierre et Marie Curie, Paris, France}
\begin{abstract}
We analyze critical behavior of superfluid density $\rho_{s}$ in
strongly disordered superconductors near superconductor-insulator
transition and compare it with the behavior of the spectral gap $\Delta$
for collective excitations. We show that in contrast to conventional
superconductors, the superconductors with preformed pairs display
unusual scaling relation $\rho_{s}\propto\Delta^{2}$ close to superconductor-insulator
transition. This relation have been reported in very recent experiments. 
\end{abstract}
\maketitle
In a disordered conductor the superconductivity is first suppressed
and then completely destroyed at high disorder. There are two mechanisms
for the suppression (and eventually the full destruction) of the superconductivity
by disorder (for recent reviews see~\cite{DG} and \cite{FIKC2010},
chapter 1). The first mechanism attributes the suppression of the
superconductivity to the increase of the Coulomb interaction that
results in the decrease of the attraction between electrons and their
eventual depairing.\cite{Finkelstein} In this mechanism the state
formed upon the destruction of the superconductor is essentially a
poor conductor. The alternative mechanism attributes superconductivity
suppression to the localization of Cooper pairs that remain intact
even when superconductivity is completely suppressed. The latter mechanism
is called bosonic and the former fermionic. The theory of the bosonic
mechanism has a long history: this scenario of the superconductor-insulator
transition was suggested long ago \cite{MaLee,KapitulnikKotliar1986,BulaSad,Ghosal2001}
but was not developed further until recently \cite{FIKY2007,FIKC2010}
when experimental data \cite{Gantm,Sh1,Sh2,SKap,Bat} indicated the
existence of a few materials that show such behavior. The bosonic
mechanism is also supported by numerical computations\cite{Trivedi2011}.

The interest to the superconductor-insulator transition without Cooper
pair destruction is both fundamental and practical. First, it provides
a perfect example of the disorder driven quantum transition in the
closed system. Second, the bosonic superconductor in the vicinity
of the transition might be ideal element for the isolation of the
coherent quantum system from the environment\textbf{\cite{DoucotIoffe2012}
}and for sensitive detectors of microwave radiation \cite{Klapwijk}.
One of the most important properties for these applications is the
value of the superfluid density of the superconductor. The goal of
this paper is to compute this quantity in the bosonic mechanism; we
also show that its behavior as the superconductivity is suppressed
distinguishes fermionic and bosonic mechanisms of the superconductivity
suppression.

Characteristic feature of the fermionic mechanism is that both the
transition temperature $T_{c}$ and the spectral gap $\Delta$ at
$T\ll T_{c}$ are suppressed simultaneously, so that their ratio is
left nearly constant in the broad range of $T_{c}$ variation. At
all disorders the superconductor remains qualitatively similar to
a conventional BCS superconductor. When the superconductivity is completely
suppressed by disorder the normal metal is formed, perhaps with a
weak tendency towards localization.

In contrast the state formed in the bosonic mechanism is qualitatively
different from both the normal metal and conventional superconductor.
In this mechanism the superconductivity disappears at the Superconductor-Insulator
transition (SIT) whilst the gap in the single electron spectrum remains
intact\cite{Sacepe2011}. Superconducting state formed in the vicinity
of SIT is therefore a pseudogapped state\cite{FIKC2010}. The best
example of such behavior is amorphous InO films in which the critical
temperature decreases by more than a factor of three while the single-particle
gap stays practically independent on $T_{c}$ but fluctuates spatially
($\Delta_{1}\approx0.45-0.6\,\mbox{meV}$) \cite{Sacepe2011}. Another
evidence of the bosonic mechanism is formation of a strong insulator
characterized by a nearly-activated dependence of resistivity $R(T)$
as observed in \cite{Sh1,Bat}.

The bosonic mechanism attributes the suppression of the superconductivity
to the competition between Anderson localization and Cooper attraction
between electrons assuming that the Coulomb interaction plays a minor
role. The key feature of this mechanism is that the superconducting
state is formed by the electrons in the \textit{localized} single-electron
eigenstates $\psi_{i}(\mathbf{r})$, with a relatively large localization
length $L_{loc}$ which depends on the proximity of the Fermi-energy
to the Anderson mobility edge $E_{c}$. The presence of a length scale
$L_{loc}$ that is not related to superconducting properties of the
conductor leads to the appearance of new energy scale in the problem
that differs from the single electron gap in a usual BCS superconductor.
Namely, attraction between two electrons that populate the same orbital
state $\psi_{i}(\mathbf{r})$, leads to the formation of a bound pair
with a binding energy 
\[
2\Delta_{1}=g\int d^{3}r\psi_{i}^{4}(\mathbf{r})\sim gL_{loc}^{-d_{2}}
\]
where $g$ is the Cooper attraction constant (with dimensionality
{[}Energy{]}$\times${[}Volume{]}), and $d_{2}\approx1.3$ is the
fractal dimension of 3D Anderson transition, see ~\cite{FIKC2010}.

\begin{figure}
\includegraphics[width=1\columnwidth]{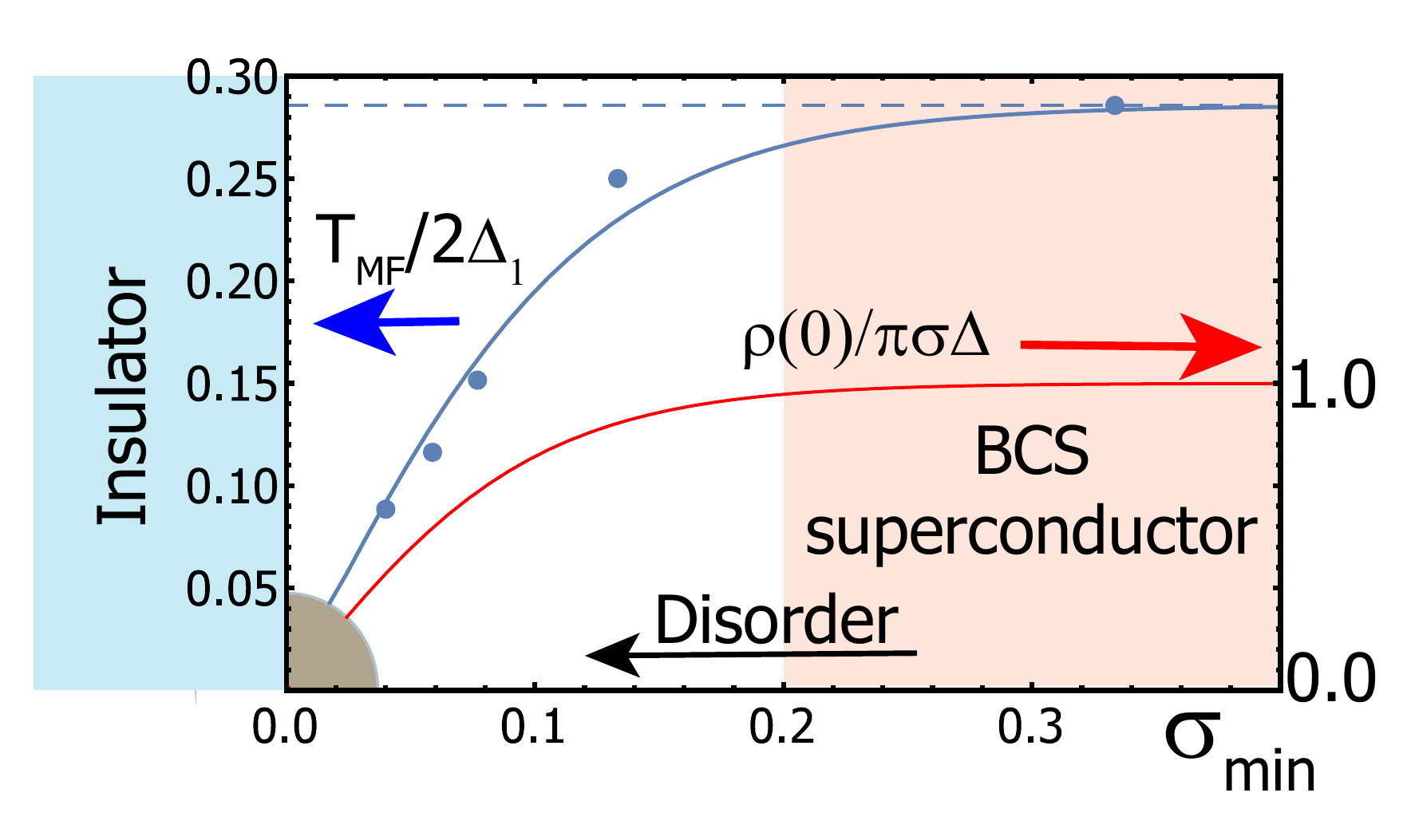}

\protect\caption{Sketch of the physical properties of the superconductors close to
the disorder driven superconductor-insulator transition. As the disorder
is increased the transition temperature drops while single particle
gap remains constant resulting in the deviations from BCS relation
$2\Delta_{1}/T_{MF}\approx3.5$ (indicated by the dashed line). At
the same time the superfluid density $\rho(0)$ at zero temperature
also drops with the respect to the predictions of the BCS theory as
shown by thin (red) line. The convenient measure of the disorder is
provided by the minimal conductance, $\sigma_{min}$ above the transition
temperature. To make the connection with the experimental situation
we have included in this plot the actual experimental data from \cite{Sacepe2011}
for InO. For these points $\sigma_{min}$ is measured in $\mbox{m}\Omega^{-1}\mbox{cm}^{-1}$.
Very close to superconductor-insulator transition superfluid density
might be controlled by non-trivial critical exponents.}
\end{figure}

The energy scale $\Delta_{1}$ discriminates against the odd (single
electron population) while even (zero or two electrons) parts of the
Hilbert space remain at low energies. As a result, single-particle
density of states (DoS) acquires a pseudogap irrespectively of the
presence of superconducting correlations. Experimentally, the appearance
of the pseudogap in the absence of long range superconducting order
was demonstrated in InO \cite{Sacepe2011} where strong suppression
of low-voltage tunneling conductance was found up to $T\geq2T_{c}$.
Similar data were reported for sufficiently disordered TiN films~\cite{BS}.
In a pseudogapped superconductor long-range correlations develop due
to coupling between localized bound pairs, and lead to formation of
another energy scale, $\Delta$, that is related to superconducting
long range order. The appearance of this energy scale was observed
recently by Andreev tunneling spectroscopy in InO$_{x}$ samples\cite{Chapelier,Dubouchet}
close to SIT with $T_{c}\sim1.5K$ . These data indicate the appearance
of the the collective energy scale $\Delta<\Delta_{1}$ that obeys
BCS-like temperature dependence while $\Delta_{1}$ is almost temperature
independent at $T<T_{c}$. Another demonstration of the presence of
some smaller energy scale that lies within the single-particle gap
is provided by optical spectroscopy of InO$_{x}$ and NbN samples
of various disorder.\cite{Sherman}

Thin superconducting films are known to possess one more important
energy scale, $\Theta$, that characterizes the dependence of the
free energy on the phase gradient: 
\[
F=\frac{1}{2}\Theta\int d^{2}(\nabla\phi)^{2}
\]
The helicity modulus $\Theta$ determines the location of the Berezinsky-Kosterlitz-Thouless
(BKT) transition~ \cite{NelsonHalperin}, at the transition $\Theta(T_{BKT})=\frac{2}{\pi}T_{BKT}$.
Energy $\Theta$ is related to the superfluid density $\rho_{s}$
which determines supercurrent in the film of thickness $d$ in presence
of a vector potential: $\mathbf{j}=-\rho_{s}\mathbf{A}/c$, namely
$\Theta=(\hbar/2e)^{2}\rho_{s}\cdot d$.

For conventional disordered superconductors there is a linear relation
(at $T\ll T_{c}$) between superfluid density $\rho_{s}$ and the
energy gap $\Delta$: 
\begin{equation}
\rho_{s}=\frac{\pi\sigma\Delta}{\hbar}\label{ns}
\end{equation}
where $\sigma$ is the normal-state resistivity~\cite{AGD}. Although
usually derived in the framework of the BCS theory this relation is
more robust because it can be traced to the optical weight conservation.
So, not surprisingly, it is valid also for the moderately disordered
InOx films with $T_{c}\approx2.5-3.5K$, as was found via the kinetic
inductance measurements~\cite{Sacepe1}. Magnetoresistance studies
of these films, see ~\cite{Sacepe2015}, as well as Andreev tunneling
spectroscopy~\cite{Dubouchet}, show expected properties of disordered
BCS superconductors, namely: no pseudogap, Andreev gap that has the
same value as the single-particle gap, finally, the destruction of
superconductivity by magnetic field leads to a normal metal state
without noticeable $R(B)$ peak.

We now demonstrate that a pseudogapped superconducting state is characterized
by a different relation between $\Theta$ and $\Delta$, leading to
a number of unique features. We start with the expression for the
full spectral weight $K^{tot}$ for frequency-dependent conductivity
as derived in Sec. 6.4 of Ref.\cite{FIKC2010}: 
\begin{equation}
K^{tot}(T)=\frac{e^{2}}{\hbar^{2}\mathcal{V}}\sum_{ij}gM_{ij}x_{ij}^{2}\frac{\Delta_{i}\Delta_{j}\tanh\beta\varepsilon_{i}\tanh\beta\varepsilon_{j}}{\varepsilon_{i}\varepsilon_{j}}.\label{K^total}
\end{equation}
Here $\beta=1/T$, $\mathcal{V}$ is the system's volume, $i,j$ enumerate
single-electron eigenfunctions those eigenvalues are $\xi_{i},\xi_{j}$,
matrix elements $M_{ij}=\int d^{3}r\psi_{i}^{2}(\mathbf{r})\psi_{j}^{2}(\mathbf{r})$,
and $\varepsilon_{i}=\sqrt{\xi_{i}^{2}+\Delta_{i}^{2}}$. The quantities
$\Delta_{i}$ are the order parameter amplitudes related to the superconducting
order parameter in a real space $\Delta(r)$ by 
\begin{equation}
\Delta(\mathbf{r})=\frac{g}{2}\sum_{i}\Delta_{i}\frac{\tanh(\beta\varepsilon_{i})}{\varepsilon_{i}}\psi_{i}^{2}(\mathbf{r})\label{Delta-r}
\end{equation}
In the mean-field approximation , the amplitudes $\Delta_{i}$ obey
self-consistency equations 
\begin{equation}
\Delta_{i}=\frac{g}{2}\sum_{j}\Delta_{j}M_{ij}\frac{\tanh(\beta\varepsilon_{j})}{\varepsilon_{j}}\label{Delta-self}
\end{equation}
Mean-field approximation of Ref.~\cite{FIKC2010} assumes that amplitudes
$\Delta_{i}$ are actually a slow functions of the single-particle
energies $\xi_{i}$, i.e. one can replace $\Delta_{i}$ by a regular
function $\Delta(\xi)$ evaluated at $\xi=\xi_{i}$; this function
has to be determined from the continuous version of Eq.(\ref{Delta-self}),
\begin{equation}
\Delta(\xi)=\frac{\lambda}{2}\int d\zeta M(\xi-\zeta)\frac{\tanh(\beta\varepsilon(\zeta))}{\varepsilon(\zeta)}\Delta(\zeta)\label{Self2}
\end{equation}
where $\lambda=g\nu_{0}$ and $\nu_{0}$ is the DoS value in the normal
state (per single spin projection), and $M(\omega)={\mathcal{V}}\overline{M_{ij}}$
with $|\xi_{i}-\xi_{j}|=\omega$.

Eq.(\ref{K^total}) was derived under the assumption that eigenfunctions
$\psi_{i,j}(\mathbf{r})$ that contribute mostly to the sum over $i,j$
are relatively well-localized, with the typical distances $x_{ij}\sim R_{0}$
between the maxima of their envelopes that are somewhat larger than
localization length $L_{loc}$. We estimate effective interaction
range as $R_{0}\sim L_{loc}\ln\frac{\delta}{\Delta}$, where $\delta_{L}=(\nu_{0}L_{loc}^{3})^{-1}$
is the level spacing within localization volume. Pseudogaped superconducting
state is realized when $\Delta<\Delta_{1}\ll\delta_{L}$, so that
$R_{0}\gtrsim L_{loc}$.

At low temperatures the optical sum rule arguments show that the major
contribution to the total spectral weight $K^{tot}$ comes from the
superconducting density, i.e $K^{tot}(0)\approx\rho_{s}(0)\equiv\rho_{s}$.\cite{FIKC2010}
The Eq.(\ref{K^total}) can be simplified by eliminating the sum over
$j$ with the help of (\ref{Delta-self}) and by replacing square
of dipole matrix elements by its average, $x_{ij}^{2}\to R_{0}^{2}/2$.
At $T=0$ we get 
\begin{eqnarray}
\rho_{s} & \approx & \frac{e^{2}R_{0}^{2}}{\hbar^{2}\mathcal{V}}\sum_{i}\frac{\Delta_{i}^{2}}{\varepsilon_{i}}=\frac{2\nu_{0}e^{2}R_{0}^{2}}{\hbar^{2}}\int_{0}^{\infty}\frac{d\xi\Delta^{2}(\xi)}{\sqrt{\xi^{2}+\Delta^{2}(\xi)}}\label{rho}\\
 & \approx & \frac{2\nu_{0}e^{2}R_{0}^{2}}{\hbar^{2}}\Delta^{2}\label{eq:rhos3}
\end{eqnarray}

We emphasize that $\Delta$ in this equation represents \textit{collective}
gap as measured by the Andreev spectroscopy\cite{Dubouchet} and in
the THz optical measurements~\cite{Sherman}, which is smaller than
single-particle gap $\Delta_{1}$.

The major difference between relations (\ref{eq:rhos3}) and (\ref{ns})
is that in the pseudogapped superconductor $\rho_{s}\sim\Delta^{2}$,
whereas in the usual case $\rho_{s}\sim\Delta$. Note that general
arguments related to optical weight conservation are not applicable~\cite{FIKC2010}
for the pseudogapped superconductor due to the presence of the second
energy scale, $\Delta_{1}$. The unusual scaling of $\rho_{s}\propto\Delta^{2}$
in Eq.(\ref{eq:rhos3}) is due to the presence of an independent spatial
scale $R_{0}$ that determines the range of the tunneling matrix elements
$M_{ij}$ between localized eigenstates $\psi_{i}(\mathbf{r})$. It
is crucial that $R_{0}$ weakly depends on $\Delta$. The counterpart
of $R_{0}$ in a usual dirty superconductor is given by its low-temperature
coherence length $\xi_{0}\approx\sqrt{\hbar D/\Delta}$. In this case
instead of $R_{0}^{2}$ one should use $\xi_{0}^{2}\propto D/\Delta$
and linear relation $\rho_{S}\propto\Delta$ is restored. Note that
quadratic scaling $\rho_{s}\propto\Delta^{2}$ is known for superfluidity
in Bose systems, see for example~\cite{Muller}. 

The prediction of unusual scaling between $\Delta$ and $\rho_{s}$
is important close to SIT where both $\Delta$ and $\rho_{s}$ are
expected to decrease strongly. Observation of the scaling $\rho_{s}\propto\Delta^{2}$
would serve as an independent demonstration of the bosonic nature
of a superconductive state.

The crucial step in the derivation of the result (\ref{eq:rhos3})
is the averaging over $\epsilon_{i}$ and $\Delta_{i}$. Close to
the transition to the insulating state the superconducting order parameter
becomes very inhomogeneous \cite{FIM2010}, which requires reexamination
of this procedure. In this regime distribution of the order parameter
becomes very broad. In the Bethe lattice approximation it is given
by 
\begin{equation}
P(\Delta)\approx\frac{\Delta_{0}^{m}}{\Delta^{1+m}}\qquad\Delta_{0}\leq\Delta\leq\Delta_{max}\label{P}
\end{equation}
Here $\Delta_{0}$ denotes the typical (most probable) value of the
local order parameter $\Delta$, the exponent $m$ is slightly less
than unity: $1-m=e\lambda\ll1$ where $e=2.718...$. Distribution
(\ref{P}) is applicable up to the upper cut-off $\Delta_{max}\sim\Delta_{1}$.

Another kind of the order parameter distribution was obtained numerically
by Lemarie et al for 2D attractive Hubbard model with strong random
potential~\cite{Lemarie2013}; it was claimed to be related to a
distribution of the Tracy-Widom type~\cite{TW} that do not have
power law tails of the form (\ref{P}), but does have a 
large dispersion.

In presence of strong statistical fluctuations of $\Delta_{i}$ the
MFA equations (\ref{Delta-self},\ref{Self2}) are not valid, thus
it is not possible to follow the route of calculations that lead to
Eq.(\ref{eq:rhos3}). Instead we note that original Eq.(\ref{K^total})
contains double sum over sites $i,j$ over large number $\sim Z\gg1$
of statistically independent terms for each $i$. Thus it is possible
to estimate this sum by 
\begin{equation}
\rho_{s}\sim\frac{\nu_{0}e^{2}R_{0}^{2}}{\hbar^{2}}\Delta_{0}^{2}\label{rhos5}
\end{equation}
where $\Delta_{0}$ is the typical (most probable) value of the local
order parameter $\Delta_{i}$, irrespectively of its specific distribution
$P(\Delta)$. Qualitatively, one expects that rare pairs of states
$i,j$ with anomalously large $\Delta_{i}\gg\Delta_{0}$ cannot contribute
considerably to macroscopic superfluid density $\rho_{s}$ since they
will be \textquotedbl{}screened\textquotedbl{} by weaker typical pairs
due to redistribution of the supercurrent density.

Notice that all quantities entering Eq.(\ref{rhos5}) except for the
interaction range $R_{0}$ are measurable. The latter can be determined
if the scaling relation (\ref{rhos5}) is observed in a broad range
of $\Delta_{0}$ and $\rho_{s}$.

Prediction (\ref{rhos5}) is in a rough qualitative agreement the
data that became available very recently \cite{Sherman}. Indeed,
total evolution of the gap as measured in this experiments by optical
spectroscopy, Fig.2c, spans about one order of magnitude between the
\textquotedbl{}crossing point\textquotedbl{}\, corresponding to the
reduced transition temperature $\tilde{T}_{c}=0.5$, and the most
disordered samples with $\tilde{T}_{c}\approx0.2$. At the same time,
superfluid density $\rho_{s}$ (extracted from the imaginary part
of the optical impedance) changes by nearly 2 orders of magnitude
in the same range of $\tilde{T}_{c}$, according to Fig.3b of the
same paper.

In the close vicinity of the quantum phase transition from superconducting
to insulating state one expects critical behavior characterized by
the exponents that might differ from the mean field result (\ref{rhos5}),
especially in low dimensional systems. For instance, the recent numerical
works~\cite{Numerics1,Numerics2} on the hard core boson model reported
behavior $\rho_{s}\propto\Delta^{a}$ with the exponent $a\approx2.5$
in the critical regime. However, applicability of 2D scaling for strongly
disordered superconducting films studied experimentally is not evident,
because many of these films are not thin enough to be considered two-dimensional.

Close to the critical point one expects large spatial flcutuations
of the order parameter. The resulting fluctuations of the superfluid
density were studied numerically in the recent paper~\cite{Numerics1}
for the 2D hard-core boson model with random potential, using Quantum
Monte Carlo method on systems with linear sizes in the interval $L=12-32$.
A broad probability distribution $\mathcal{P}(\ln\rho_{s})$ was found
both in superfluid and insulating phases; however, on superfluid side
of the transition, the width of this distribution diminishes with
$L$, whereas an opposite tendency is found for the insulating state.
These results indicate that in a macroscopic system \textit{dc} superfluid
density is a self-averaging quantity in the superconducting state.
Analytical calculation of the corresponding distribution $\mathcal{P}(\ln\rho_{s})$
in the framework of the theory developed here is an interesting problem
which we leave for future studies.

The spatial fluctuations of $\rho_{s}$ are probably observable in
the low frequency measurements. Usually $\rho_{s}$ is measured via
\textit{ac} kinetic inductance $\mathcal{L}\propto1/\rho_{s}$, see
for example~\cite{Armitage,Yazdani}. Nonzero measurement frequency
$\omega$ determines the length scale $l_{\omega}=\sqrt{D/\omega}$
where coherent transport takes place. With a typical diffusion contant
for very strongy disordered superconductors $D\sim0.1-1cm^{2}/s$,
the length $l_{\omega}$ ranges from  $0.02 - 0.5$ micron for high-frequency
measurments~\cite{Armitage} to  $ 10 - 100 $ micron for low-frequency ones~\cite{Yazdani}.
In the latter case it might be possible to detect spatial fluctuations
of $\rho_{s}$ with a scanning technique. 

Sharp drop near SIT of the helicity modulus $\Theta=(\hbar^{2}/4e^{2})\rho_{s}\cdot d$
according to (\ref{rhos5}) leads to an enhancement of both thermal
and quantum phase fluctuations, and to suppression of both $T_{c}$
and critical magnetic field $H_{c}$ with respect to the mean-field estimates, see also Ref.~\cite{Seibold}. 

In conclusion we have shown that the bosonic mechanism of the superconductor-insulator
transition implies a nearly quadratic scaling of the superfluid density
with the superconducting order parameter, the prediction that can
be verified experimentally. This resulting very low values of the
superfluid density would make materials close to the transition extremely
useful for the applications.

This research was supported by the Russian Science Foundation grant
\# 14-42-00044. We are grateful to B. Sacepe for useful discussions.

\end{document}